\documentclass[%
 reprint,
 superscriptaddress,
 amsmath,amssymb,
 aps,
 prl,
]{revtex4-2}

\usepackage{amsmath, amssymb}
\usepackage[arrowdel]{physics}
\usepackage[version=4]{mhchem}
\usepackage{graphicx}
\usepackage[]{float}
\usepackage[]{booktabs}
\usepackage{svg}
\usepackage{appendix}
\usepackage[margin=0.5in]{geometry}

\usepackage{braket}
\usepackage{dcolumn}
\usepackage{bm}

\usepackage{mathtools}
\usepackage[
]{color}
\usepackage[colorlinks=true, citecolor=blue, linkcolor=blue, urlcolor=blue]{hyperref}
\usepackage{xr-hyper}
\usepackage[printonlyused]{acronym}
\usepackage{layouts}
\usepackage{url}

\makeatletter
\newcommand*{\addFileDependency}[1]{
  \typeout{(#1)}
  \@addtofilelist{#1}
  \IfFileExists{#1}{}{\typeout{No file #1.}}
}
\makeatother

\setcitestyle{super}

\svgpath{../figures/}
\graphicspath{{../figures/}}

\begin{document}

\title{Casimir Stabilization of Fluctuating Electronic Nematic Order}

\author{Ola Carlsson}
\affiliation{Institute for Theoretical Physics, ETH Zürich, Zürich 8093, Switzerland}
\affiliation{Department of Physics and Arnold Sommerfeld Center for Theoretical Physics (ASC),
Ludwig-Maximilians-Universität München, Theresienstr. 37, D-80333 München, Germany}

\author{Sambuddha Chattopadhyay}
\affiliation{Institute for Theoretical Physics, ETH Zürich, Zürich 8093, Switzerland}
\affiliation{Lyman Laboratory, Department of Physics, Harvard University, Cambridge MA, USA}

\author{Jonathan B. Curtis}
\affiliation{Institute for Theoretical Physics, ETH Zürich, Zürich 8093, Switzerland}

\author{Frieder Lindel}
\affiliation{Institute for Theoretical Physics, ETH Zürich, Zürich 8093, Switzerland}
\affiliation{Quantum Center, ETH Zürich, Zürich 8093, Switzerland}

\author{Lorenzo Graziotto}
\affiliation{Institute of Quantum Electronics, ETH Zürich, Zürich 8093, Switzerland}
\affiliation{Quantum Center, ETH Zürich, Zürich 8093, Switzerland}

\author{Jérôme Faist}
\affiliation{Institute of Quantum Electronics, ETH Zürich, Zürich 8093, Switzerland}
\affiliation{Quantum Center, ETH Zürich, Zürich 8093, Switzerland}

\author{Eugene Demler}
\affiliation{Institute for Theoretical Physics, ETH Zürich, Zürich 8093, Switzerland}

\begin{abstract}
Vacuum cavity control of quantum materials is the engineering of quantum materials systems through electromagnetic zero-point fluctuations.
In this work we articulate a generic mechanism for vacuum optical control of correlated electronic order: Casimir control, where the zero-point energy of the electromagnetic continuum, the Casimir energy, depends on the properties of the material system. 
To assess the experimental viability of this mechanism we focus on the Casimir stabilization of fluctuating nematic order. In nematic Fermi liquids, different orientations of the electronic order are often energetically degenerate. Thus, while local domains of fixed orientation may form, thermal disordering inhibits long range order. 
By engineering the electromagnetic environment of the electronic system, however, we show that the Casimir energy can be used as a tool to preferentially stabilize particular orientations of the nematic order. 
As a concrete example, we examine the interplay between a birefringent crystal---which sources an anisotropic electromagnetic environment---and a quantum Hall stripe system, an archetypal nematic Fermi fluid. We show that for experimentally feasible setups, the anisotropy induced by the orientation dependent Casimir energy can be $10^4$ times larger than other mechanisms known to stabilize quantum Hall stripes. This finding convincingly implies that our setting may be realized with currently available experimental technology. Having demonstrated that the Casimir energy can be used to stabilize fluctuating nematic order, we close by discussing the implications for recent terahertz cavity experiments on quantum Hall stripes, as well as pave the road towards broader Casimir control of competing correlated electronic phases.
\end{abstract}

\maketitle

\begin{figure}[h]
        \centering
        \includegraphics[width=1\linewidth]{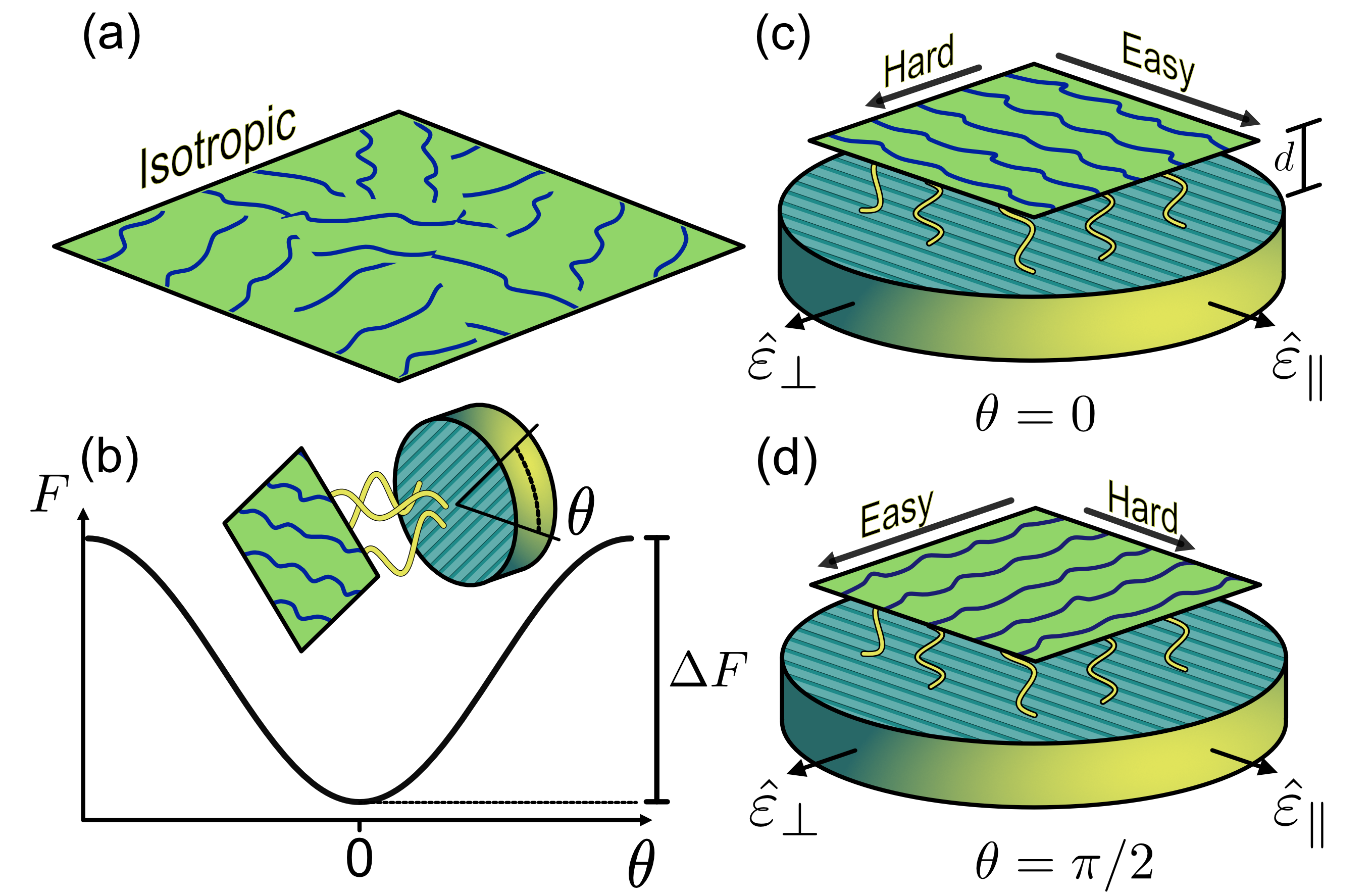}
        \caption{\textit{Schematic picture of Casimir stabilization of the stripe phase.} 
        With no symmetry breaking field (a), stripes form thermally disordered domains, and the macroscopic sample is isotropic. In the presence of a parallel birefringent plate (optic axis \(\varepsilon_\| > \varepsilon_\perp\)) (b), the vacuum electromagnetic modes give stripes a twist angle \(\theta\) dependent free energy, aligning them at an energy scale \(\Delta F\) per unit area.
        Conductivity easy axis of the stripes aligned with high dielectric function \(\hat{\varepsilon}_\|\) of the controlling plate (c) is energetically preferred to conductivity hard axis aligned with \(\hat{\varepsilon}_\|\) (d).}\label{fig:overiew}
\end{figure}
\section{Introduction}

The optical control of quantum materials is one of the central projects of contemporary condensed matter physics\cite{basov2017towards}. While progress in laser technology over the past decades has enabled non-thermal forms of control \textit{out-of-equilibrium} \cite{de_la_torre_colloquium_2021}
, a complementary direction seeks to control correlated electronic systems using ultra-strong light-matter interactions arising from the confinement of electromagnetic fields using cavities\cite{Schlawin.2022}. This mode of control seeks to leverage the quantum fluctuations of the electromagnetic vacuum to control material properties in equilibrium. The idea of using the \textit{vacuum} field of the cavity for control has been explored theoretically in diverse contexts ranging from ferroelectricity~\cite{ashida2020quantum, curtis2023local}
to superconductivity~\cite{schlawin2019cavity, PhysRevLett.122.167002}
to ferromagnetism~\cite{roman2021photon}. Experiments have successfully uncovered additional mechanisms for cavity material control beyond these proposals~\cite{appugliese_breakdown_2022, enkner_tunable_2025, jarc_cavity-mediated_2023}, spurring theoretical efforts to conceptualize experimentally grounded regimes of cavity engineering.
As an example, recent experiments have shown that the presence of terahertz (THz) split ring resonators destroys integer quantum Hall effect signatures for nearby Hall bars~\cite{appugliese_breakdown_2022}. Crucially this breakdown persists over a large range of magnetic fields, suggesting that the anomalous transport properties in the cavity rely on aspects of the cavity-matter system exogenous to the strong coupling between the cyclotron resonance in the 2DES and the fundamental mode of the cavity.  This observation is consistent with recent theoretical work, in which systematically treating the full continuum of electromagnetic modes has been found essential when controlling low-energy excitation properties~\cite{sanchez2024general, pantazopoulos2024electrostatic}. 
Indeed modifying ground state energies---the target of many vacuum cavity engineering efforts---has proven problematic in single or few-mode approaches~\cite{kotov_casimir-lifshitz_2025}, suggesting that conceptualizing approaches to vacuum cavity control requires carefully reconciling contributions from all modes in the electromagnetic environment. 

A prominent physical manifestation of the quantum fluctuations of the electromagnetic vacuum is the Casimir effect\cite{casimir1948attraction}. It is a vacuum force arising between reflecting bodies due to the Casimir energy, a shift in the electromagnetic vacuum’s zero-point energy due to the imposed boundary conditions. A non-resonant phenomenon that relies on the entire electromagnetic continuum, the Casimir energy between two bodies is contingent on various context specific details, in particular the optical properties of the materials at hand. Traditionally, the Casimir effect is considered a force acting on macroscopic bodies, the result of the Casimir energy's dependence on geometric parameters. We consider instead the implication of the energy's dependence on the optical properties of one object, a ``force'' on the electronic degrees of freedom that determines the reflectivity of the body. By tailoring the electromagnetic environment of an electronic system with an engineered electromagnetic structure, one can exploit the specificity of Casimir physics, using it as a tool to steer among \textit{optically distinct}, energetically competing phases, tilting the energetic balance towards the phase that lowers the Casimir energy: 
We call this \textit{Casimir control}, a mechanism through its mode continuum nature suitable for electromagnetic engineering in non-resonant regimes.

By virtue of the vacuum nature of this mechanism, one might fear that the energy scales involved should be far too small to impact electron systems. To establish a sense of scale it is fruitful to consider the following estimate. The (appropriately renormalized\footnote{The electromagnetic vacuum has a divergent zero-point energy both in free space and in the presence of perturbing structures. The (physical) Casimir energy is the \textit{difference} between the energy shift in the presence of perturbing structures and the divergent contribution from the vacuum. For example, for the perfectly conducting plates the Casimir energy is the energy computed for inter-plate separation $d$  minus the result when $d \to \infty$.}) areal Casimir energy density between two perfectly conducting plates at a habitually experimentally achievable distance of $10 \text{ nm}$ apart is $2.7\times 10^{11}$ ${\rm eV cm^{-2}}$. Per contrast, the intense energetic competition between strongly correlated two-dimensional electronic phases with \textit{distinct optical properties} often occurs at the $\mu {\rm eV}$ ($\sim 10 {\text{ mK}}$) per particle scale. Typical electron densities for the same systems are ($\sim 1 \times 10^{11} \text{ cm}^{-2}$ ), amounting to areal energy densities of \(\sim 1\times 10^5\) \(\rm eV cm^{-2}\), six orders of magnitude smaller than the Casimir energy scale. Our estimate suggests that if one can harvest even a minute fraction of the Casimir energy between the engineered electromagnetic structure and the electronic system, then the Casimir energy can be used to control properties of the electronic system.

In this work, we seek to examine the feasibility of this program in a setting that is particularly amenable to Casimir control: \textit{electronic nematic order}\cite{fradkin_nematic_2010}. Examples of electronic nematic order include forms of charge density wave order---found in quantum Hall systems and bilayer ruthenate \ce{Sr3Ru2O7} under a strong magnetic field---and cuprate and iron-based high temperature superconductors. Nematic order manifests in systems where the rotational symmetry of the underlying lattice is spontaneously broken by the \textit{electronic} degrees of freedom without significant structural distortion. Nematic phases typically display large transport and optical anisotropies arising from different symmetry broken orientations (e.g., in nematic charge density order, the orientation corresponds to direction of charge ordering). Different orientations are often energetically (near) degenerate and thermal disordering of locally formed ordered domains precludes long range ordering. The anisotropic optical properties that arise from the distinct orientations, however, provide a handle for Casimir control, allowing for the Casimir \textit{stabilization} of a preferred orientation that minimizes the Casimir energy. 

In order to situate our analysis within a concrete setting, we consider a specific form of electronic nematic order: quantum Hall stripes. This nematic occurs in quantum Hall systems---two-dimensional electron systems (2DES) with a perpendicular magnetic field---at moderately large field strengths and ultralow temperatures. In half-filled high Landau levels (LLs), an interaction-driven instability produces the stripes, a charge–density-wave like pattern, with an ordering length on the order of the cyclotron radius\cite{koulakov1996charge, fogler2002stripe}. 
Different orientations of the stripe order are energetically degenerate, and displays in general thermal disordering into local domains.
The presence of \textit{tiny} anisotropies---estimated to be on the order of $1$ ${\rm mK}$ per particle\cite{cooper_investigation_2001}---are known to break the orientational degeneracy of the stripes, leading to an onset of macroscopic charge ordering.

Our choice to focus on quantum Hall stripes is particularly motivated by recent experiments wherein the low-temperature electronic transport of a quantum Hall system interacting with the vacuum field of a metamaterial cavity (a terahertz slot antenna resonator) was investigated\cite{graziotto_cavity_2025}. At high half-integer fillings of the quantum Hall system, a giant anisotropy of the resistance---reaching a $50$ fold discrepancy between the longitudinal resistivities---was observed when the sample temperature was lowered below $\sim 200 {\text{ mK}}$. Remarkably, this effect is induced by the cavity as transport is isotropic outside of the resonator. The results of these experiments were attributed---by several authors of this manuscript---to arise from the cavity aligning pre-existing orientationally fluctuating quantum Hall stripes into long-range order, leading to dramatically anisotropic longitudinal transport. Despite the explanatory power of such an interpretation, a precise quantitative understanding of how the cavity preferentially favors the alignment of the stripes is still an open question. Notably, the qualitative aspects of the effect are experimentally robust to differences in density, filling factor, and design of metamaterial cavity, implying that the orientational aligning mechanism at play is non-resonant. This suggests that a departure from the quantum optics inspired single-cavity mode approach \cite{Lu} is necessary, to be replaced by approaches commonly used to treat cavity-mediated dispersion forces in nanophotonics\cite{BuhmannWelsch2007PQE,pantazopoulos2024electrostatic,sanchez2024general}.

Inspired by the experiments which demonstrated the vacuum control of quantum Hall stripes through \textit{cavity} means\cite{graziotto_cavity_2025}, in this work we consider a birefringent plate, \ce{BaTiO3}, placed in close proximity to a quantum Hall system. 
Even though such a plate does not form a separate cavity on its own, hosting discrete modes, it can influence the electron system through the Casimir control mechanism.
Working within the quantum Hall stripe regime, we demonstrate how this leads to macroscopic stripe stabilization. Note that in this scenario, a cavity is rather formed \textit{between} the electron system and birefringent plate, elevating the quantum system studied to an essential component of the setup geometry.
Our proposed physical setting allows us to examine the feasibility of Casimir control within an analytically tractable setting that is additionally amenable to experiments.  We first show how the Casimir contribution to the free energy of our system depends on stripe order orientation, showing that in certain regimes it can far exceed the energies required to align stripes. We then examine the dependence of this contribution on various experimentally relevant parameters, showing that the distance between the birefringent plate and the quantum Hall system is the most salient aspect for control. Our feasibility study evinces that a fully non-resonant Casimir control may enable formation of macroscopic electron nematic order, within a formalism that accounts for the contributions from the full electromagnetic continuum.

\section{Results}
\subsection{System description}
We consider a quantum Hall system ---a 2DES hosted in an ultra-high mobility semiconductor heterostructure with an out-of-plane magnetic field---in the high LL regime (\(\nu > 4\)), where \(\nu = 2\pi n \hbar /e B\) is the filling factor of the quantum Hall system, i.e.\ the number of occupied LLs. If the filling factor is additionally at half-integer values \(\nu = N + 1/2\) (with \(N\) integer), the ultra-low temperature (\(\sim 100\) mK) ground state is the quantum Hall stripe phase of charge-density wave type order, manifesting as enormous magnetotransport anisotropies\cite{fogler_ground_1996, fradkin_liquid-crystal_1999, lilly_evidence_1999}.
Intuitively, the magnetotransport anisotropies arise from the fact that transport is easy along the edge of a density modulation (stripe) while transport across the stripes is hard, requiring scattering between stripes, see figure~\ref{fig:overiew}. 
Macroscopic observation of this anisotropy however implies a broken rotational symmetry as the orientation of the stripes fixes the easy and hard axes. While there is full rotational symmetry within Hartree-Fock theory~\cite{fogler_ground_1996}, in the presence of a weak orienting field the rotational symmetry is broken at low temperatures. 
Estimates of the scale of this symmetry breaking field 
place it at $\sim 1 \text{ mK}$ per particle~\cite{cooper_investigation_2001}, inferred from experiments which showed that quantum Hall stripes can be reoriented by using an in-plane magnetic field~\cite{pan_strongly_1999}.  In absence of such a field, domains of stripe order are thermally scrambled with respect to their orientation, as illustrated in figure~\ref{fig:overiew} (a), leading to fully isotropic macroscopic transport properties.

We propose non-resonant vacuum-mode \textit{Casimir control} of this system to play the role of a symmetry-breaking field, leading to giant macroscopic transport anisotropies, similar to what was uncovered in recent experiments\cite{graziotto_cavity_2025}. This is achieved through the introduction of a birefringent, i.e. optically anisotropic~\cite{hecht_optics_2017}, crystal in close proximity to the 2DES, whose 
boundary conditions on the vacuum electromagnetic modes act to steer the properties of the electronic phase in the 2DES. For concreteness we consider this to be a plate of \ce{BaTiO3} placed in parallel configuration to the plane of the quantum Hall stripe system, at a distance \(d\) much shorter than the width of the birefringent plate, with its optic axis parallel to its surface.  The presence of the plate modifies the electromagnetic normal modes in the volume between the two objects, leading to a Casimir energy contribution. 

We focus our attention on the anisotropy introduced by the birefringence of the plate. For every thermally fluctuating domain, the conductivity tensor will exhibit an easy-axis along the extent of the stripes. This easy-axis extends an angle \(\theta\) away from the optic axis of the birefringent plate, see figure~\ref{fig:overiew} (b). We will show in the following, as an exact result, how the vacuum free energy density \(F\) of the electromagnetic modes depends also on this twist angle \(\theta\). This is conceptually related to the Casimir 
\textit{torque}\cite{barash_moment_1978,philbin_alternative_2008,somers_measurement_2018,spreng_recent_2022}, previously discussed for setups of two birefringent plates and other geometries\cite{rodrigues_vacuum-induced_2006, lindel_inducing_2018}.  Minimizing the free energy with respect to \(\theta\) allows for an energy gain on the order of \(\Delta F\) per unit area, defined as the largest difference in the free energy with respect to \(\theta\). This thus adds incentive to align all fluctuating domains with respect to the birefringent plate, forming a macroscopic stripe phase with magnetotransport anisotropy.
In addition the direction of the stabilized macroscopic stripe phase is imposed by the birefringent optic axis, through the angle \(\theta\) that minimizes the free energy. We will see that this ideal angle is close to \(0\), such that the 2DES easy axis is aligned with the birefringent plate optic axis, see figure~\ref{fig:overiew}(c-d). 
\subsection{Casimir control from scattering theory}

The free energy contribution from the electromagnetic continuum between an anisotropic 2DES and a birefringent \ce{BaTiO3} plate is divergent even at zero temperature, and must be renormalized by subtracting the energy content of free space electromagnetic modes in the same volume. The scattering formalism of the Casimir effect takes this renormalization into account, and non-perturbatively calculates the renormalized free energy per unit area~\cite{lambrecht_casimir_2006, rahi_scattering_2009}:
\begin{align}\label{eq:omegasc}
F(\theta) &= \frac{\hbar}{2\pi} \int d\omega_m d \vb{k}_\perp \ln
\det \qty(1 - e^{-2\kappa d}R^{\alpha\beta}_e (\theta) R^{\beta\gamma}_b)
.\end{align}
Here \(\omega = i\omega_m\) are imaginary frequencies (\(\omega_m\) is a real quantity), \(\vb{k}_\perp\) is the wave vector component parallel to the surfaces, \(\kappa = \sqrt{\omega_m^2 /c^2 +
\vb{k}_\perp^2} \) is the (evanescent) wave number in the direction normal to the
surfaces, \(d\) is the distance between the planes, and \(R_e\), \(R_b\) are the reflection matrices of the 2DES and birefringent plate respectively. 
The reflection matrices are additionally functions of frequency \(i\omega_m\) and momentum \(\vb{k}_\perp\). The usage of imaginary frequencies deserves particular mention, and implies that 
the reflection matrices appearing in equation~\eqref{eq:omegasc} do not directly represent physical scattering events of waves oscillating at frequency \(\omega\). Instead the reflection matrices for real frequencies, describing physical scattering, have been \textit{analytically continued} with respect to frequency into the complex plane, and evaluated on the imaginary axis.
The integral over imaginary frequencies is equal to a similar integral over real frequencies and physical scattering events, allowing us to choose between either formulation. The benefits of the imaginary frequencies, as well as further details on equation~\eqref{eq:omegasc}, are discussed in Methods.

Crucially the reflection matrix of the electron system \(R_e\) depends on its conductivity, which sets the boundary conditions respected by reflected electromagnetic waves. Since this conductivity depends on the direction \(\theta\) of anisotropic order, so does the reflection matrix, \(R_e = R_e(\theta)\), conferring the total free energy the same twist angle dependence, \(F = F(\theta)\). Through this mechanism it is the characteristic property, transport anisotropy, of nematic Fermi liquids which allows for Casimir stabilization, independent of model details.

The thick slab of birefringent crystal, aiming to control the behavior of the 2DES, can be modeled as a half-plane of dielectric function  \(\varepsilon^{\alpha\beta} = \mathrm{diag}(\varepsilon_\|(i\omega_m), \varepsilon_\perp(i\omega_m),\varepsilon_{\perp}(i\omega_m))\)
where the optic axis \(\hat{\varepsilon}_\|\) and plane perpendicular to it has dielectric functions \(\varepsilon_\|(i\omega_m)\) and \(\varepsilon_\perp(i\omega_m)\), respectively. For the considered \ce{BaTiO3} we have \(\varepsilon_\|(i\omega_m) > \varepsilon_\perp(i\omega_m)\) for all relevant frequencies. The dielectric functions are evaluated on the imaginary axis, that is the directly measurable real frequency values have to be analytically continued. Presently for the \ce{BaTiO3} we use the analytic continuation of two-oscillator models, see Methods.
For a pedagogical introduction to the interpretation and calculation of imaginary frequency dielectric functions, see \cite{hough_calculation_1980}.

To be concrete, we now focus on the specific example of the anisotropic Fermi system: the stripe nematic Fermi liquid that emerges at high Landau levels in  quantum Hall systems.
We phenomenologically capture the essential characteristics---highly anisotropic longitudinal conductivities \(\sigma_1\), \(\sigma_2\), as well as the Hall conductivity \(\sigma_h\)---using a modified Drude model with anisotropic effective masses, see Methods 
for details:
\begin{align}
        &\sigma(i\omega_m) = \nonumber\\&\frac{R_K^{-1} \nu \Omega_c \tau}{\Omega_c^2
        \tau^2 + (1 + \omega_m \tau)^2} \mqty(
                \sqrt{\lambda} (1 + \omega_m  \tau) & \Omega_c \tau \\
                -\Omega_c \tau & \sqrt{\lambda}^{-1}(1 + \omega_m \tau))
                \label{eq:cond}
.\end{align}
This conductivity is characterized by the cyclotron frequency \(\Omega_c = eB /m^*\) where \(m^*\) is the zero-field effective mass, the anisotropy ratio \(\lambda = \sigma_1 /\sigma_2\), and a phenomenological relaxation time \(\tau\). The conductivity is naturally given in units of the inverse von Klitzing constant \(R_K^{-1} = e^2 /2\pi\hbar\).

From the dielectric function tensor of the birefringent plate and the conductivity tensor of the stripes, given in eq.~\eqref{eq:cond}, the reflection matrices \(R_e\), \(R_b\) can be found as they depend on momentum \(\vb{k}_\perp\), imaginary frequency \(i\omega_m\), and twist angle \(\theta\), see Methods for explicit expressions. This allows for exact, non-perturbative evaluation of the free energy \(F(\theta)\), enabling precise qualitative and quantitative statements about Casimir control in this geometry. 

While the Casimir energy is in principle temperature dependent, sizable variations occur on a temperature scale \(T_d = \hbar c /k_B d\) set by the interplane distance \(d\)~\cite{spreng_thermal_2025}. At sub-micron distances \(d < 1\) µm this scale is bigger than \(\sim 10^3\) K, whereas stripes only emerge at the local level at $\sim 1$ ${\rm K}$.  Accordingly, we consider the zero-temperature limit of the Casimir energy.

\subsection{Casimir control alignment}
\begin{figure}[h]
        \centering
        \includesvg[width=1\linewidth]{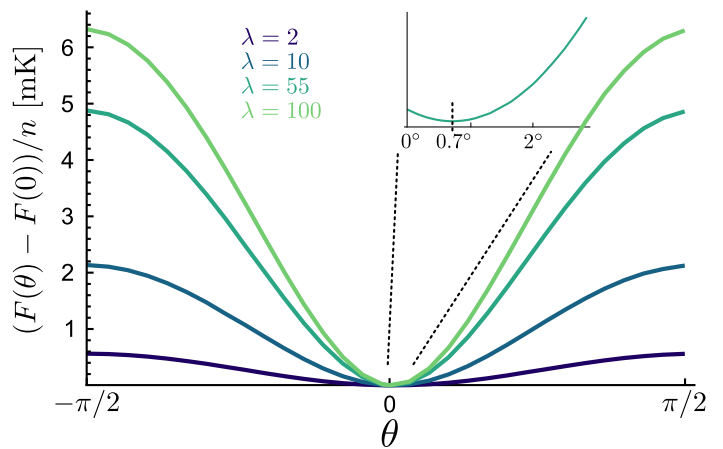}
        \caption{\textit{Free energy per particle \(F(\theta)/n\), as a function of twist angle \(\theta\).}
                Values are given relative to the configuration with 2DES easy
                axis \(\hat{e}\) parallel with the birefringent plate optic axis
                \(\hat{\varepsilon}_\|\) (\(\theta = 0\)). A few anisotropy ratios of the stripe phase \(\lambda = \sigma_1 /\sigma_2\) are shown. Parameters are \(d = 100\) nm, \(\nu = 15/2\), \(\tau = 1\) ns and \(n = 2.9\times 10^{11}\) cm\(^{-2}\). Due to the presence of Hall conductivity, the free energy minimum is shifted away from the symmetry point \(\theta = 0\) (inset).}\label{fig:lambdaplot}
\end{figure}

In order to assess the experimental feasibility of Casimir control we evaluate numerically the free energy per particle \(F(\theta)/n\), with \(n\) the electron density, in Figure~\ref{fig:lambdaplot} for different twist angles \(\theta\) and anisotropy ratios \(\lambda = \sigma_1 /\sigma_2\). 

For stripe phases such anisotropy ratios on the order of \(10\) are common, with the highest reliable reported experimental value at \(55\)~\cite{sammon_resistivity_2019}.  The alignment of the electronic phase domain such that easy-axis conductivity is parallel with the \ce{BaTiO3} optic axis has a lower free energy and is therefore preferred. This is equivalent to the alignment of axes of high dielectric function, and agrees with previous findings on the Casimir torque~\cite{spreng_recent_2022}.

The energy scale of the stabilization reaches a few mK per particle, for anisotropy ratios \(\lambda \ge 10\), exceeding the \(\sim 1\) mK threshold reported by earlier studies of the in-plane magnetic field experiments\cite{ivar, cooper_investigation_2001}. In passing, we note that, remarkably, this energetically preferred alignment is not exact (i.e., not $0^\circ$), visible in the inset of figure~\ref{fig:lambdaplot}, but rather deviates with \(\sim 0.7^\circ\), a deviation that is induced by the finite Hall conductivity, see Appendix. This deviation arising from Hall conductivity is, to our knowledge, a new result for Casimir physics. 
\subsection{Dependence of the Stabilization Energy on Experimental Parameters}
\begin{figure}[h]
        \centering
        \includesvg[width=1\linewidth]{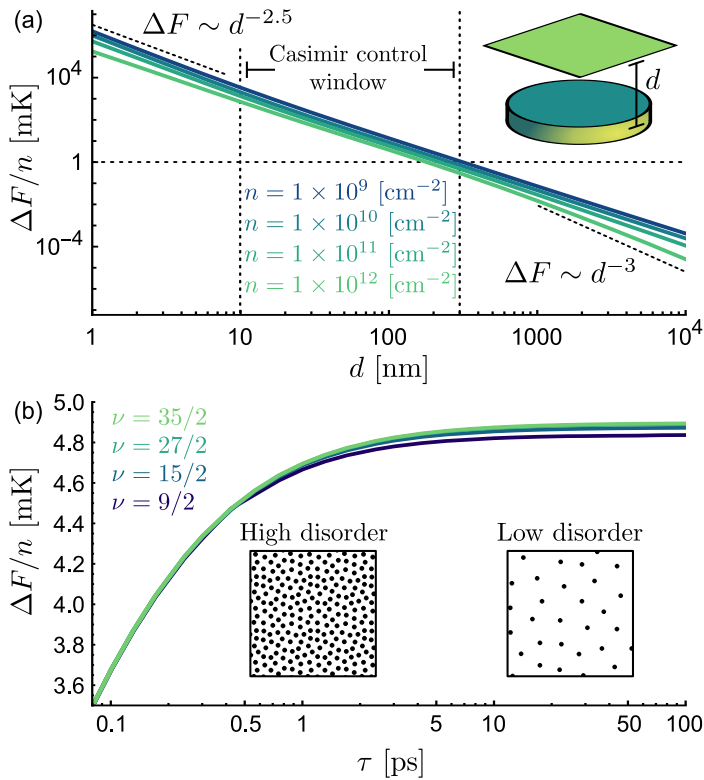}
        \caption{\textit{Stabilization energy per particle \(\Delta F /n\) as a
                        function of interplane distance \(d\) and electron density \(n\) (a), and as a function of Drude scattering time \(\tau\) and filling factor \(\nu\) (b).} The anisotropy ratio is the highest experimental value \(\lambda = 55\), remaining parameters are \(d = 100\) nm, \(\nu = 15/2\), \(\tau = 1\) ns and \(n = 2.9\times 10^{11}\) cm\(^{-2}\), unless varied. The horizontal line marks the onset of relevant stabilization magnitudes \(\Delta F /n = 1\) mK. Vertical lines show the window of relevant distances. Dashed slanted lines show the long-distance behavior \(\Delta F \sim d^{-3}\) above the range plotted, and the non-retarded behavior in the limit of vanishing density \(n\), \(\Delta F \sim d^{-2.5}\).
                }\label{fig:dntaunuplot}
\end{figure}
Now we plot in figure~\ref{fig:dntaunuplot} the dependence 
of the Casimir control energy scale \(\Delta F\) on geometric and 2DES parameters. In contrast to the theoretically 
proposed resonant cavity QED control paradigms---where a 
single photon mode is carefully engineered to maximize its 
impact on a target system---the Casimir control is non-resonant and depends on the properties of \textit{all} 
electromagnetic modes. Thus, to investigate how a 
particular experimental geometry controls the phase of the 
electronic system, a careful accounting of the entire 
electromagnetic continuum must be undertaken. Within our 
analysis we vary aspects of the controlling setup---the 
distance between the birifrengent plate and the 2DES---and properties of the electronic system including the scattering time (modeling the relevance of disorder) in the finite-frequency conductivity model, the density, and the filling factor.  We find that while the Casimir stabilization energy is relatively invariant to properties of the electronic system, the distance between the plate and the 2DES can radically change the energy scale of the induced energy. Crucially, we find a large, experimentally viable range of distances where the Casimir energy is sufficient to stabilize fluctuating stripe order.  

In Fig.~\ref{fig:dntaunuplot} (a) we examine the distance dependence, for a large range of electron densities \(n\). We find that the Casimir stabilization remains relevant, i.e.\ larger than \(\sim 1\)~mK per particle, from the sub-micron range of a few hundred nanometers and below.  Additionally the energy scale \(\Delta F\) increases significantly as the distance is lowered, reaching an astounding \(1-10\) K per particle at experimentally feasible distances of \(\sim 10\) nm, far beyond the $1$ ${\rm mK}$ scale required to stabilize a particular stripe orientation. The electron density has a much weaker impact than the orders of magnitude which would be expected just from the per particle normalization. This arises from the fact that the conductivity is favored by larger density, with higher conductivities in turn benefiting the Casimir control due to stronger boundary conditions imposed on the electromagnetic modes. That higher densities benefit the Casimir stabilization, reducing the dependence of per particle energy scales on \(n\), implies that observations of Casimir control will not hinge crucially on the electron densities achievable.

We also comment briefly on the asymptotic behavior in the limit of small and large distances. The characteristic length scales determining the onset of both regimes are set by the frequency dependence of the optical responses involved\cite{BuhmannWelsch2007PQE}, which in our present case amounts to distances far above \(30\text{ cm}\) or far below \(100 \text{ nm}\). 
For the distance window amenable to experiment then, the Casimir physics lies between these asymptotic regimes, where algebraic behavior \(\Delta F\sim d^{-3}\) or \(\Delta F\sim d^{-2.5}\) is expected for long and short distances, respectively. Interestingly, the non-standard short-distance exponent of \(-2.5\) is due to the two-dimensional nature of the 2DES, and only applies to the Drude model without cyclotron resonance~\cite{khusnutdinov_casimir_2019}, i.e.\ here in the limit \(n \to 0\).

In Fig.~\ref{fig:dntaunuplot} (b), we plot the dependence of our suggested control mechanism against variations of additional stripe phase properties: the filling factor \(\nu\) for fixed $n$, as well as the Drude scattering time \(\tau\). 
Disorder in the 2DES enters our model as the Drude relaxation time \(\tau\). We find that it impacts the stabilization magnitude weakly. In addition, in the limit \(\tau \to 0\), the stabilization is independent of the filling factor \(\nu\). This we understand as the conductivity being completely dominated by disorder scattering, with a formal return to regular metallic Drude behavior \(\sigma \sim n \tau\), independent of cyclotron resonance effects that carry information on the filling factor. In the limit of low disorder, the 2DES conductivity is affected only by the cyclotron resonance, with the stabilization magnitude plateauing as a function of \(\tau\). The dependence on the cyclotron frequency remains weak however, and in combination with the strongly sublinear dependence of the free energy on the conductivity, we find an astonishingly small impact on the stabilization. Filling factors in the range displaying stripe phases \(9/2 \le \nu \le 35/2\) barely affect the Casimir stabilization at all, signifying universal control strength across a large regime of quantum stripe Hall phases. This shows us additionally that the veracity of the relaxation time assumption \(\tau \approx 1\) ns, taken in previous plots, is not very important. Stripe phases occur in very pure high mobility \ce{AlGaAs}/\ce{GaAs} heterostructures~\cite{sammon_resistivity_2019}, where the corresponding scattering times are large (\(\sim 1 \) ns), and we generally expect variations in disorder to remain in the plateau limit \(\tau \to \infty\). Only scattering times on the low sub-picosecond scale see any modification to this Casimir control strength, where the order of magnitude of the stabilization remains unaffected.
\section{Discussion \& Outlook}
In our work we have examined the experimental feasibility of using the Casimir effect to stabilize fluctuating pre-formed nematic Fermi liquid phases, in particular of quantum Hall stripes. Our investigation of a particular control set-up---using an optically birefringent \ce{BaTiO3} plate---reveals that the Casimir stabilization mechanism can reach stabilization energy scales up to \(10^4\) times larger than previously known mechanisms for macroscopic stripe alignment\cite{stanescu_finite-temperature_2000, cooper_investigation_2001}. We stress that even though our system partially serves to demonstrate the viability of the Casimir control mechanism at large, in light of our analysis, the birefringent plate control of a quantum Hall system provides a simple and compelling experimental set up for the vacuum electromagnetic control of correlated quantum order. 
In particular our setup carries the benefit of admitting even quantitative estimates of optical control strengths. By introducing an in-plane magnetic field, much like carried out in the studies of stabilization through inherent crystal anisotropies~\cite{pan_strongly_1999, cooper_investigation_2001}, a tunable orientation selecting field competing with the preferred direction of the Casimir control can be introduced. We believe that measurements of the critical in-plane field, at which the preferred conductivity easy-axis of the birefringent plate is overwhelmed by the easy-axis preferred by the magnetic field, can allow precise estimates of the Casimir control energy scale.
We also mention that another promising realization of Casimir stabilization lies in utilizing two-dimensional anisotropic metamaterials for the purpose of engineering the electromagnetic environment. For example, arrays of metallic ellipses or wires deposited on the 2DES, with a spacer layer in between controlling the distance, effect the same stabilization mechanism as that of the birefringent plate, due to the optical anisotropy of the arrays. This might be favorable to experimental considerations, when compared to the birefringent plate.
Our work should also encourage investigations---theoretical and experimental---of the Casimir stabilization of other, phenomenologically similar electronic nematic orders: specifically \ce{Sr3Ru2O7} in a large magnetic field but also, perhaps more speculatively, cuprate and iron-based high-temperature superconductors\cite{fradkin_nematic_2010}.

The promising energy scale of our results lends support to related nematic order stabilization mechanisms proposed in recent experiments on quantum Hall systems in terahertz slot antenna resonators \cite{graziotto_cavity_2025}. We note however, that an important distinction between the experimental setting explored in this work and the experiments performed in terahertz resonators is that the anisotropy of the electromagnetic environment in our case arises from material anisotropies intrinsic to \ce{BaTiO3}, whereas in the case of the terahertz slot antenna resonator, anisotropies arise from the geometry of a cavity. 
In fact, the setup considered in the present work does not control the electron system by means of a separate resonator, hosting sharp modes in the absence of the 2DES, at all, further separating the possible conceptual and quantitative treatments of the birefringent plate and terahertz cavity experiments. Indeed, by considering this geometry, we broaden the scope of electromagnetic vacuum control outside of the commonly considered vacuum cavity regime. 
This approach emphasizes which modifications to the electromagnetic environment are most relevant for effective control, stripping away superfluous aspects of resonator design that do not impact experiments in vacuum settings.
The generality of the Casimir control mechanism nevertheless allows application of our framework to geometries utilizing distinct cavities, although careful computational examination with an \textit{ab initio} model of the specific cavity is needed in order to solidify quantitative statements about experiments such as~\cite{graziotto_cavity_2025}.

We next underscore the crucial mode continuum nature of the Casimir control mechanism, wherein there is no explicit selection of a few modes that dominate the physics, in line with existing experiments which permit an equilibrium understanding\cite{appugliese_breakdown_2022, enkner_tunable_2025, graziotto_cavity_2025}. Indeed, the mode continuum contributions within our work underlie the large energy scales reported herein, in line with a steady current of works which similarly find that a careful accounting of multi-mode contributions (e.g., \cite{riolo_tuning_2025, eckhardt_surface-mediated_2024}) lead to much more sizable effects once experimental parameters are considered than single-mode treatments\cite{Zeno}. It is crucial, however, that to accurately treat these mode continuum systems, it is necessary to properly renormalize divergent quantities, lest even qualitative predictions such as signs of energy shifts be erroneous. The Casimir control formalism provides a prescription for this procedure through the subtraction of free-space zero-point contributions, 
a prescription
previously verified experimentally from a host of successful Casimir force measurements. 

In our work we show the promise of Casimir control to stabilize orientationally fluctuating nematic Fermi liquids. We close by noting that we believe that our promising results in this---admittedly favorable---context suggest that Casimir control to more broadly \textit{select} among competing phases in strongly correlated electronic systems is a promising frontier for vacuum electromagnetic control. An ideal platform for this endeavor would be Moiré systems, which share the phenomenology of low-temperature quantum Hall systems, particularly the fact that these systems are also two-dimensional and harbor various energetically close competing phases arising from interactions in a quenched kinetic energy setting\cite{mak_semiconductor_2022, Andrei2020}.  

On a technical level this effort would proceed as follows. First the competing correlated phases of interest would have to be characterized by their optical conductivities \(\sigma^{\alpha\beta}_{l}(\vb{k}, i\omega_m)\), where the index \(l\) runs over different phases, and the corresponding reflection matrices \(R^{\alpha\beta}_l(\vb{k}_\perp, i\omega_m)\) found. We note in passing that the non-local nature of the optical conductivity might matter for set-ups where the perturbing structures reach distances at the scale of the mean free path of the electron. For larger distances---indeed as relevant for the majority of experimental setups---a theory for the local optical conductivity should prove sufficient, since the momentum dependence of reflection matrices is dominated rather by contributions from Maxwell's equations.
Next, unique features of the reflectivity of the desired phase  should be suitably matched by an intentional choice of the nearby object to perform the Casimir control. A characteristic frequency could for example be matched by a similarly resonating dielectric, maximizing the overlap of high reflection frequency ranges, or a highly non-local conductivity could be matched by an appropriate antenna or metamaterial, reflecting selectively the most relevant momenta. Last, the Casimir free energy (Eq.~\eqref{eq:omegasc}) can be directly computed for all competing phases in the presence of the same reflecting object. This is interpreted as a unique effective energy shift to every phase, rearranging the free energy landscape and potentially leading to selection of a new ground state. 

\section{Methods}
\label{sec:methods}
\label{app:refl}
The free energy of equation~\ref{eq:omegasc} is found as an integral over imaginary frequencies \(\omega = i\omega_m\) (\(\omega_m\) is a real quantity), such that the integrand does not directly describe physical scattering events, but has rather been \textit{analytically continued} into the complex plane of frequencies.
In principle integration of the same object could instead be carried out over real frequencies, lending direct interpretation of the integrand as the relevant density of states, or physical scattering events, for the electromagnetic continuum. This freedom of choice is a consequence of the properties of contour integration, as well as of the causality of the system implying domains of analyticity. We choose to work with imaginary frequency rather than real frequency, as the integrand then becomes a monotonically decreasing function, and the absence of the oscillatory behavior present for real frequencies allows for much more stable numerical evaluation. The integrand is even suppressed exponentially for large \(\omega_m\), and only a small range of frequencies demand evaluation.

This conciseness of the imaginary frequency formulation can additionally be considered a consequence of the non-resonant nature of the Casimir energy\cite{kotov_casimir-lifshitz_2025}: every considered imaginary frequency \(i\omega_m\) in principle contains information on \textit{all} real frequencies \(\omega\), and we may describe contributions from the entire electromagnetic continuum in the small range of imaginary frequencies integrated over.
See ref.~\onlinecite{rodriguez_virtual_2007} for a more in-depth discussion on the analytic continuation and choice of imaginary frequencies. 

The Casimir energy depends directly on the reflection properties of the birefringent plate and the 2DES. These can be found from matching boundary conditions of electromagnetic waves scattering off of each object in isolation. Here we summarize the optical properties and reflection coefficients of the two bodies, and show how they depend on the twist angle \(\theta\), transferring this dependence to the free energy. We work in SI units, and with dielectric functions given relative to the vacuum permittivity \(\varepsilon_0\).

\begin{figure}[h]
        \centering
    \includegraphics[width=0.9\linewidth]{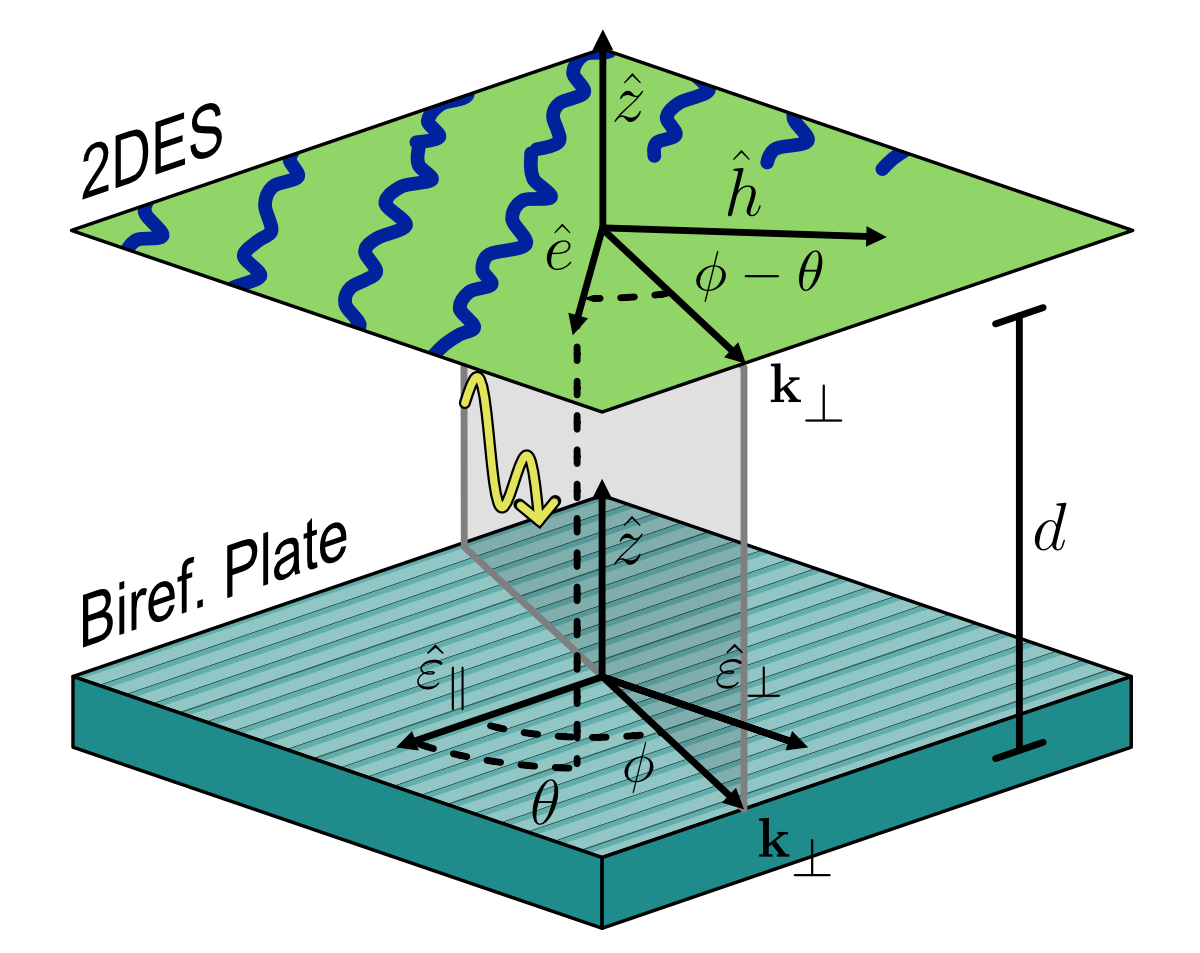}
        \caption{\textit{Coordinate basis for the 2DES and plate twisted at an angle \(\theta\).} A photon (yellow arrow) of in-plane momentum \(\vb{k}_\perp\) 
        makes an angle \(\phi\) with the optic axis \(\hat{\varepsilon}_\|\) of the birefringent plate. The vectors \(\hat{\varepsilon}_\perp\) and \(\hat{z}\) complete the basis. The same photon makes an angle \(\phi - \theta\) with the easy axis \(\hat{e}\) of the electron system.}\label{fig:twisted1}
\end{figure}

\subsection{Birefringent plate details}
We begin with finding the optical properties of the birefringent plate.
Let \((\hat{\varepsilon}_\|\text{, } \hat{\varepsilon}_\perp, \hat{z})\) be a basis of the principal axes of the crystal, such that \(\hat{z}\) is normal to the interfaces. The dielectric function tensor of the birefringent plate is then in this basis given by \(\varepsilon^{\alpha\beta} = 
\mathrm{diag}(\varepsilon_\|, \varepsilon_\perp, \varepsilon_\perp)\)
, where we remind ourselves that all components are frequency dependent: \(\varepsilon_\| = \varepsilon_\|(i\omega_m)\text{, }\varepsilon_\perp = \varepsilon_\perp(i\omega_m)\).
The dielectric properties are sufficiently captured
by a two-oscillator model~\cite{munday_torque_2005, meurk_direct_1997}
:
\begin{align*}
        \varepsilon_{\| /\perp}(i\omega_m) &= 1 + \frac{C_{\text{IR}}}{1 + (\frac{\omega_m}{\omega_{\text{IR}}})^2}
        + \frac{C_{\text{UV}}}{1 + (\frac{\omega_m}{\omega_{\text{UV}}})^2}
,\end{align*}
where \(C_{\text{IR}}\), \(C_{\text{UV}}\), are the absorption strengths and \(\omega_{\text{IR}}\),
\(\omega_{\text{UV}}\), are the characteristic absorption frequencies in the infrared and
ultraviolet ranges, respectively. The relevant parameters for \ce{BaTiO3} are summarized in
Table~\ref{tab:BaTiO3}.
\begin{table}[H]
        \caption{\textit{Optical parameters for} \ce{BaTiO3}~\cite{munday_torque_2005, meurk_direct_1997}.
        }
        \label{tab:BaTiO3}
        \centering
\begin{tabular}{@{}l|ll@{}}
                        & \(\varepsilon_\|\)      & \(\varepsilon_\perp\)   \\ \midrule
\(C_{\text{IR}}\)              & \(3595\)                & \(145.0\)               \\
\(C_{\text{UV}}\)              & \(4.128\)               & \(4.064\)               \\
\(\omega_{\text{IR}}\) (rad/s) & \(0.850\times 10^{14}\) & \(0.850\times 10^{14}\) \\
\(\omega_{\text{UV}}\) (rad/s) & \(0.841\times 10^{16}\) & \(0.896\times 10^{16}\)
\end{tabular}
\end{table}

We now turn to the dependence of the reflection coefficients on these dielectric functions. The reflection coefficients are also functions of both photon transverse momentum magnitude \(k_\perp\), parameterizing the angle of incidence, and of momentum direction. The latter follows from the anisotropy of the reflecting surface, and is eventually translated into a dependence on twist angle.
If \(\phi\) is the azimuthal angle of the photon in-plane momentum \(\vb{k}_{\perp}\) from \(\hat{\varepsilon}_\|\)
towards \(\hat{\varepsilon}_\perp\), the reflection coefficients for the birefringent plate in an \(s\)-wave, \(p\)-wave
basis
can be written~\cite{spreng_recent_2022, lekner_reflection_1991}
:
        \begin{align}
                r^b_{ss} &= r_D^{-1}(\sin^2(\phi) \tilde{\alpha}_- \gamma_+ +
                \cos^2
                (\phi)
                \alpha_- \nu_+)\nonumber\\
                r^b_{pp} &= - r_D^{-1} ( \sin^2(\phi) \tilde{\alpha}_+ \gamma_- -
                \cos^2(\phi)
                \alpha_+ \nu_-)\nonumber\\
                r^b_{ps} &= r^b_{sp} = r_D^{-1}\omega_m
                \varepsilon_{\perp}\kappa_{\perp}\kappa(\kappa_\| -
                \kappa_\perp)\sin(2\phi)\nonumber\\
                r_D &= \sin^2(\phi) \tilde{\alpha}_+\gamma_+ + \cos^2(\phi)
                        \alpha_+ \nu_+\nonumber\\
                \alpha_{\pm} &= \kappa \pm \kappa_\perp\nonumber\\
                \tilde{\alpha}_\pm &= \kappa \pm \kappa_{\|}\nonumber\\
                \nu_\pm &= \kappa_{\perp}^3 \pm \varepsilon_\perp
                \kappa\kappa_\perp\kappa_\|\nonumber\\
                \gamma_\pm &= \varepsilon_\perp \omega_m^2 
                \qty(\varepsilon_\perp \kappa \pm \kappa_\perp)\nonumber
        ,\end{align}
        where the ordinary and extraordinary \(\hat{z}\) wave numbers are
        \begin{align*}
                \kappa_\perp &= \sqrt{\varepsilon_\perp \omega_m^2 + k_\perp^2} \\
                \kappa_\| &= \sqrt{\varepsilon_\| \omega_m^2 + k_\perp^2 +
                (\varepsilon_\|  / \varepsilon_\perp - 1) k_\perp^2 \cos^2 (\phi)} 
        .\end{align*}
        See Figure~\ref{fig:twisted1} for a sketch of the system with
        coordinates and angles.

\subsection{Electron system details}
Quantum Hall stripes manifest in experiment as enormous magnetotransport anisotropies for Hall bars with sufficiently low disorder.
These anisotropies occur in the longitudinal conductivities, which are finite for this quantum Hall system since stripes occur at half-integer filling factors \(\nu\), i.e.\ our discussion is that of a compressible electron state.
For the purposes of the Casimir stabilization we model
the 2DES as a two-dimensional layer of finite anisotropic in-plane two-dimensional
conductivity. In a coordinate system \((\hat{e}, \hat{h},
\hat{z})\) of conductivity easy axis, hard axis, and out-of-plane
normal, respectively, we can write the full conductivity
tensor in the form \(\smqty(\sigma_1 & \sigma_h & 0 \\ -\sigma_h & \sigma_2 & 0\\ 0 & 0 & 0)\). Also here all components are frequency dependent, which we model by a modified Drude approach, capturing the essential transport anisotropy and Hall conductivity of the stripe phase.

We write down the equation of motion for a Drude model in two dimensions, with a magnetic field \(\vb{B}\) in the out of plane direction, and with an anisotropic mass tensor \(\vb{\tilde{m}} = m^* \mathrm{diag}(\lambda^{-1 /2}, \lambda^{1/ 2})\):
          \begin{align*}
        \vb{\tilde{m}} \frac{d}{dt}\vb{v} &= -e \vb{E} + e \vb{v} \times \vb{B} - \vb{\tilde{m}} \tau^{-1}\vb{v}
.\end{align*}
Here \(\vb{v}\) is the velocity of the charge carriers, \(-e\) is the electron charge, \(\vb{E}\) an applied electric field, \(m^*\) the effective mass of the zero-field 2DES, and \(\tau\) some phenomenological scattering time.
Solving for the conductivity we find:
\begin{align*}
        \sigma(i\omega_m) &= \frac{R_K^{-1} \nu \Omega_c \tau}{\Omega_c^2
        \tau^2 + (1 + \omega_m \tau)^2} \mqty(
                 (1 + \omega_m  \tau)\sqrt{\lambda} & \Omega_c \tau \\
                -\Omega_c \tau & (1 + \omega_m \tau)/\sqrt{\lambda})
.\end{align*}
Here \(\Omega_c = eB/m^*\) is the cyclotron frequency, \(\nu = 2\pi n \hbar/eB\) the filling factor of Landau levels, and \(R_K^{-1} = e^2 /2\pi\hbar\) the inverse von Klitzing constant.

One consequence of this functional form is that the highest longitudinal conductivity is not found at \(\omega_m = 0\), but rather at \(\omega_m = \Omega_c - \tau^{-1}\), where it takes on the values:
\begin{align*}
        \sigma_{max} &= R_K^{-1}\frac{\nu}{2} \times \begin{cases}
               \sqrt{\lambda},  &\text{Easy axis}\\
               \sqrt{\lambda}^{-1}, &\text{Hard axis}
        \end{cases}
.\end{align*}
We note then that for \(\Omega_c > \tau^{-1}\) the cyclotron frequency and relaxation time do not impact the maximum value of the conductivity, but that this is set entirely by filling factor \(\nu\) and anisotropy ratio \(\lambda\).

We now consider the reflection matrix of the 2DES, as it depends on this conductivity.  
The reduced dimensionality of the plane modifies the form of these reflection coefficients when compared to the birefringent plate, such that birefringent plate dielectric function and stripe conductivity do not play equal roles.
Matching free space waves on each side of the plane with
discontinuities imposed by the finite two-dimensional conductivity yields
the reflection matrix, which can be
        written~\cite{antezza_casimir-polder_2020}:
        \begin{align}
                R^{\alpha\beta} &= - \frac{2\omega_m^2 \sigma^{\alpha\beta} +
                        2k_\perp^\alpha 
                k^\gamma\sigma^{\gamma\beta} + \delta^{\alpha\beta} \kappa \omega_m
                \det \vb{\sigma}}{2\omega_m^2 \trace{\vb{\sigma}} + 2k^\gamma k_\perp^\nu
                \sigma^{\gamma\nu} + \kappa \omega_m(4 + \det \vb{\sigma})}
        .\nonumber\end{align}
        Solving for the reflection coefficients in a \(s\)- and \(p\)-wave
        basis, it is convenient to write down the conductivity in the corresponding in-plane directions, that is:
        \begin{align*}
                \sigma^{\alpha\beta} 
                &= \sigma_0 \mqty(1 & 0 \\
                0 & 1)
               + \frac{\sigma_\delta}{2}\mqty(\cos(2\varphi) &
                -\sin(2\varphi)  \\
                -\sin(2\varphi) & -\cos(2\varphi)
                ) + \sigma_h \mqty(0 & 1\\ -1 & 0)
                \\ &= \mqty(\sigma_{pp} & \sigma_{sp} + \sigma_h\\
                \sigma_{sp} - \sigma_h & \sigma_{ss})
        ,\end{align*}
        where \(\sigma_0 = \frac{\sigma_1 + \sigma_2}{2}\) and \(\sigma_\delta = \sigma_1 - \sigma_2\) are the mean and difference of
        the easy and hard conductivities, respectively.
        We define \(\varphi\) as the angle of \(\vb{k}_\perp\) 
        from \(\hat{e}\) to \(\hat{h}\). We may then write down the reflection coefficients:
        \begin{align*}
                r^e_{ss} &= - g^{-1} (\sigma_{ss}(2\omega_m^2 + \kappa\omega_m
                \sigma_{pp})+ \kappa \omega_m(\sigma_h^2 - \sigma_{sp}^2))\\
                r^e_{pp} &= - g^{-1} (\sigma_{pp}(2\kappa^2 + \kappa\omega_m
                \sigma_{ss}) + \kappa\omega_m(\sigma_h^2 - \sigma_{sp}^2))\\
                r^e_{ps} &=  -2 g^{-1}\omega_m^2 (\sigma_{sp} - \sigma_h)\\
                r^e_{sp} &= -2 g^{-1} \kappa^2 (\sigma_{sp} + \sigma_h)\\
                g &= (2\kappa + \omega_m\sigma_{ss})(2\omega_m +
                \kappa\sigma_{pp}) + \kappa\omega_m (\sigma_h^2 - \sigma_{sp}^2)
        \end{align*}
      where now \(\sigma_{ij}\) depend on the angle \(\varphi\).  
        
We summarize the reflection coefficients of both bodies in the reflection matrices:
\begin{align*}
        R_e^{\alpha\beta} &= \mqty(r^e_{ss} & r^e_{ps}\\
        r^e_{sp} & r^e_{pp})\\
        R_b^{\alpha\beta} &= \mqty(r^b_{ss} & r^b_{ps}\\
        r^b_{sp} & r^b_{pp})
.\end{align*}
Considering the easy axis \(\hat{e}\) of the 2DES to be offset by an angle
\(\theta\) from the birefringent plate optic axis \(\hat{\varepsilon}_\|\), see figure~\ref{fig:twisted1}, a single photon momentum \(\vb{k}_\perp\) is associated with two different azimuthal angles for the two reflecting surfaces. The birefringent slab reflection matrix
\(R^{\alpha\beta}_b\) is for one momentum evaluated at an angle \(\phi\), while the reflection matrix for the 2DES \(R^{\alpha\beta}_e\) is evaluated  at an angle
\(\varphi = \phi - \theta\).

Parameterizing the integration over \(\vb{k}_\perp\) by its azimuth \(\phi\) and its magnitude \(k_\perp\), we may calculate the areal density of the Casimir free energy contribution from equation~\ref{eq:omegasc}:
\begin{align}
   F &= \frac{\hbar}{(2\pi)} \int_0^\infty d\omega_m \int_0^{\infty} d k_\perp k_\perp \int_0^{2\pi} d\phi   \nonumber\\ \times \bigg[
    & \ln
\det \qty(1 - e^{-2\kappa d}R^{\alpha\beta}_a(\phi - \theta) R^{\beta\gamma}_b(\phi))\bigg]
,\nonumber\end{align}
where we keep in mind that the reflection matrices depend on momentum \(\vb{k}_\perp\) and imaginary frequency \(\omega_m\) both explicitly, and implicitly through the frequency dependence of the material responses \(\varepsilon_\|\), \(\varepsilon_\perp\), and \(\sigma^{\alpha\beta}\).

\section{Code availability}
The code that supports the plots within this paper are available from the corresponding author upon reasonable request.

\bibliography{references}{}

\section{Acknowledgments}
The authors thank Misha Fogler, Dima Basov, Atac Imamoglu, Marios H. Michael, Johannes Feist, Carlos J. Sánchez Martínez, and Oleg Kotov for fruitful discussions. L.G. and J.F. acknowledges funding from the Swiss National Science Foundation (SNF Project
No. 10000397). 
O.C., S.C., J.B.C., and E.D. acknowledge support from the SNSF project 200021\_212899 and the Swiss State Secretariat for Education, Research and Innovation (SERI) under contract number UeM019-1.
This work was supported by the Quantum Center Research Fellowship and the Dr Alfred and Flora Spälti Fonds. This project has received funding from the European Research Council (ERC) under the European Union’s Horizon 2020 research and innovation program (Grant Agreement no 948141) — ERC Starting Grant SimUcQuam.

\section{Author contributions}
S.C., J.B.C., L.G., E.D., and J.F. developed the conceptual approaches of this work.  O.C., S.C, J.B.C., F.L., and E.D. designed the research and interpreted the data. O.C. performed calculations and numerical simulations. E.D. supervised the work. All authors contributed to the manuscript.

\section{Competing interests}
The authors declare no competing interests.

\section{Correspondence}
Correspondence and requests for materials should be addressed to Ola Carlsson or Eugene Demler.

\appendix
\section{Symmetry breaking by Hall conductivities}\label{app:bfield}
\begin{figure}[h]
        \centering
        \includesvg[width=1\linewidth]{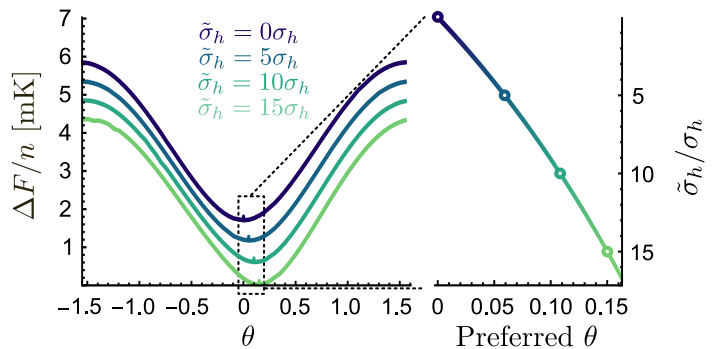}
        \caption{\textit{Free energy per particle \(F(\theta) /n\) as a
                        function of twist angle \(\theta\) for increased Hall conductivities \(\tilde{\sigma}_h\).} Here the anisotropy ratio is at the highest experimental value \(\lambda = 55\), remaining parameters are \(d = 100\) nm, \(\nu = 15/2\), \(\tau = 1\) ns and \(n = 2.9\times 10^{11}\) cm\(^{-2}\).
                        Finite Hall conductivity increases the Casimir stabilization
                        slightly,
                        as well as shifts the preferred twist angle (marked) away from
                        the symmetry point \(\theta =0\) (blowup).
                }\label{fig:hallplot}
\end{figure}

The dependence of the Casimir energy on angles of rotation corresponds directly to the previously studied Casimir torque~\cite{spreng_recent_2022}.
Typical settings for Casimir torque effects contain sufficient symmetry to restrict global minima of the free energy to the twist angles \(\theta = 0\), \(\pi /2\), and therefore the preferred configuration of two objects will always be with their optic axes either aligned or perpendicular to each other.

For the present case of the stripe quantum Hall system however, introduction of finite Hall conductivities breaks the relevant symmetry, allowing the preferred twist angle to vary continuously. We explore this in figure~\ref{fig:hallplot}, where we artificially set the Hall conductivity to a different value \(\tilde{\sigma}_h\) than that imposed by our anisotropic Drude model, and observe a sublinear shift of the energetically preferred angle when increasing the Hall conductivity.

Additionally, we below derive the constraint usually restricting the optimal angle to \(0\) or \(\pi/2\), and show how the Hall conductivity, or equivalently a finite magnetic field, violates the assumptions required for this argument.

Considering two objects with conduction easy-axes, or otherwise optic axes that are bi-directional, the system is invariant to twist rotations of \(\pi\). Whether two optical axes are parallel or anti-parallel, for example, does not distinguish two configurations. Therefore, \(F(\theta) = F(\theta + \pi)\), and we can always consider \(0\le \theta <\pi\).

\begin{figure}[h]
        \centering
        \includesvg[width=1\linewidth]{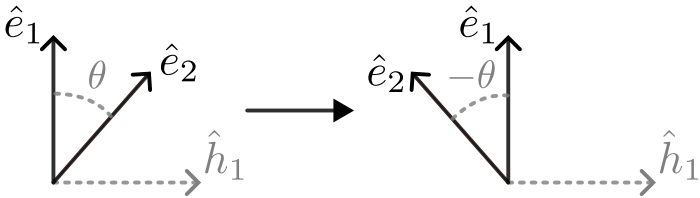}
        \caption{\textit{Reflection symmetry of the Casimir torque.} The system is symmetric for reflections in the plane spanned by e.g.\ easy axis conductivity \(\hat{e}_1\) and out of plane \(\hat{z}\) (not shown). The hard axis \(\hat{h}_1\) here completes the right-handed orthonormal basis \((\hat{e}_1, \hat{z}, \hat{h}_1)\), and is therefore not transformed, while the easy axis of the second object \(\hat{e}_2\) now forms the angle \(-\theta\) after the reflection.}\label{fig:reflection}
\end{figure}

For much the same reason, the system is also invariant under reflection in the plane spanned by \(\hat{e}_1\) and \(\hat{z}\), where \(\hat{e}_1\) is the easy-axis direction of one object (or any other in-plane optical symmetry axis). This reflection reverses the twist angle, and so \(F(\theta) = F(-\theta)\), see figure~\ref{fig:reflection}.
Any angle \(\theta_0\) that minimizes the free energy therefore implies another minimum at \(\theta = \pi - \theta_0\): \(F(\theta_0) = F(\pi - \theta_0)\). If the minimum is to be global, these two angles must coincide modulo \(\pi\), and we find that \(\theta_0 = 0\) or \(\pi /2\).

Now consider a finite Hall contribution to the conductivity:
\begin{align}
                \sigma^{\alpha\beta} &\sim \mqty(0 & \sigma_h & 0\\
                -\sigma_h & 0 & 0\\
                0 & 0 & 0) \label{eq:2deshallcond}
        .\end{align}
This is not invariant under the previously described reflection, which is equivalent to conjugation with \(\smqty(1 & 0 & 0\\ 0 & -1 & 0 \\ 0 & 0 & 1)\)
, and we might therefore have that \(F(\theta) \neq F(-\theta)\).
We might also consider that Hall conductivity is the result of finite magnetic field. Being a pseudo-vector, this magnetic field always changes sign under a reflection, and the same conclusion can be drawn.
As a result the global minima of the free energy may be situated at any angle between \(0\) and \(\pi\).
\end{document}